\documentclass[printer]{aa}  

\usepackage[varg]{txfonts}

\usepackage{natbib}

\usepackage{xcolor}
\usepackage{amsmath}

\newcommand{\Msun}{\mbox{M$_{\odot}$} }

\newcommand {\AU} {\, \mathrm{AU}}
\newcommand {\pc} {\, \mathrm{pc}}

\newcommand {\nbody} {\textsc{\mbox{nbody6 }}} 

\begin{document}

\title{Does the mass distribution in discs influence encounter-induced losses in young
    star clusters?}
\titlerunning{Influence of disc-mass distribution on disc frequency in 
    star clusters}
\author{Manuel Steinhausen
  \and Susanne Pfalzner}
\authorrunning{Steinhausen
  \and Pfalzner
}
\institute{Max-Planck-Institut f\"ur Radioastronomie, Auf dem H\"ugel 69, 53121 Bonn, Germany\\
  \email{mstein@mpifr.de}
}

\date{ }

\abstract{
One mechanism for the external destruction of protoplanetary discs in young dense clusters is tidal disruption during the flyby of another cluster member. The degree of mass loss in such an encounter  depends, among other parameters, on the distribution of the material within the disc. 
Previous work showed that this is especially so in encounters that truncate large parts of the outer disc. The expectation is that the number of completely destroyed discs in a cluster depends also on the mass distribution within the discs.
}
{ Here we test this hypothesis by determining the influence of encounters on the disc fraction and average disc mass in clusters of various stellar densities for different mass distributions in the discs.
}
{ This is done by performing \nbody simulation of a variaty of cluster environments, where we track the encounter dynamics and determine the mass loss due to these encounters for different disc-mass distributions.
}
{ We find that although the disc mass distribution has a significant impact on the disc losses for specific star-disc encounters, the overall disc frequency generally remains rather unaffected. The reason is that in single encounters the dependence on the mass distribution is strongest if both stars have very different masses. Such encounters are rather infrequent in sparse clusters. In dense clusters such encounters are more common, however, here the disc frequency is largely determined  by encounters between low-mass stars such that the overall disc frequency does not change significantly. 
}
{For tidal disrution the disc destruction in clusters is fairly independent of the actual distribution of the material in the disc. The all determining factor remains the cluster density.}

\maketitle


\section{Introduction}
\label{sec:intro}

Young stars are initially surrounded by a circumstellar disc. Observations show that with time the circumstellar discs in young clusters become depleted of gas and dust and eventually
disappear. It is currently unclear which physical mechanism dominates the evolutionary disc destruction processes. Among the great
variety of effects are internal processes such as viscous torques \citep[e.g.][]{1987ARA&A..25...23S}, turbulent effects
\citep{2003ApJ...582..869K}, and magnetic fields \citep{2002ApJ...573..749B}, but as well external disc destruction processes like
photoevaporation \citep{2001MNRAS.325..449S, 2001MNRAS.328..485C, 2003ApJ...582..893M, 2004RMxAC..22...38J, 2005MNRAS.358..283A,
  2006MNRAS.369..229A, 2008ApJ...688..398E, 2009ApJ...699L..35D, 2009ApJ...690.1539G} and tidal interactions
\citep{1993ApJ...408..337H, 1993MNRAS.261..190C, 1994ApJ...424..292O, 1995ApJ...455..252H, 1996MNRAS.278..303H,
  1997MNRAS.287..148H, 1997MNRAS.290..490L, 1998MNRAS.300.1189B, 2004ApJ...602..356P, 2005ApJ...629..526P, 2006ApJ...653..437M,
  2008A&A...487..671K}.

In this study we investigate the early embedded phase of stellar cluster development with the scope to understand the general influence of gravitational interactions on the disc frequencies. In contrast to previous investigations 
\citep{1993MNRAS.261..190C, 1996MNRAS.278..303H, 1998MNRAS.300.1189B, 2004ApJ...602..356P,
  2005ApJ...629..526P,2006ApJ...642.1140O, 2006ApJ...653..437M, 2008A&A...487..671K} we concentrate on the influence of the mass distributions in the disc on the resulting disc frequency.

Due to the temperature and dust grain size distribution in discs, infra-red observations can only resolve the inner areas ($< 10 \AU$).  Alternatively, observations in the sub-millimeter range are limited to the outerskirts ($50 \AU$). So no continuous observational spatial coverage of entire discs with high resolution exist to date. Nevertheless, by fitting resolved millimeter
continuum or line emission data with parametric disc structure models \citep[e.g.][]{1996ApJ...464L.169M, 1997ApJ...489..917L} a wide variety of different surface densities profiles have been derived. An other method has been the combination with broadband spectral energy distributions (SEDs) \citep{2000ApJ...534L.101W, 2001ApJ...554.1087T,
  2002ApJ...566.1124A, 2002ApJ...581..357K, 2007ApJ...659..705A}. Those studies have profoundly shaped our knowledge of disc structures, however, all have fundamentally been limited by the low angular resolution of available data. Thus to date there is no conclusive knowledge about the typical surface density of protoplanetary discs and how it develops with time.  

Most previous numerical studies of star-disc encounters have used a theoretically motivated \mbox{$r^{-1}$-dependent} disc-mass
distribution \citep{1996MNRAS.278..303H, 1997MNRAS.287..148H, 2004ApJ...602..356P, 2006ApJ...642.1140O, 2006ApJ...653..437M,
  2007A&A...462..193P}. Recently \citet{2012A&A...538A..10S} investigated for a wide parameter space the relative disc-mass loss
in encounters with special focus on the dependence on the mass distribution in the disc and found differences of up to $40 \, \%$
between the different initial density distributions for the same type of encounter. The aim of this study is to investigate
whether this means a comparable difference in the encounter-induced disc frequency in clusters.

The likelihood of encounters is a function of the stellar density which varies between different clusters but as well within the considered clusters. To first order the encounter
  frequency, $\eta$, depends on the stellar number density, $n_{\mathrm{star}}$, as $\eta \propto n_{\mathrm{star}}$
  \citep{2008gady.book.....B}. This approximation is valid for systems with equal-mass stars undergoing two-body encounters. For
  systems with unequal stellar masses effects like gravitational focusing become important  \citep{2008gady.book.....B} and for very high stellar densities the
  two-body approach breaks down \citep{2013A&A...555A.135P}. The cluster-average densities span a wide range from sparse
clusters like Taurus with densities of $1-10$ stars per $\pc^{3}$ \citep{2009ApJ...703..399L} to very massive and compact clusters
like the Arches cluster with a core density of several $10^5$ stars per $\pc^{3}$ \citep{1999ApJ...525..750F}. Such dense stellar
environments lead to strong gravitational interactions between the cluster members and affect as well the protoplanetary discs.

The mass (and angular momentum) losses due to gravitational interactions in clusters as a function of stellar densities has been
investigated in several studies \citep{2004ApJ...602..356P, 2006ApJ...642.1140O, 2006ApJ...641..504A, 2007A&A...462..193P}. In
particular, \citet{2006ApJ...642.1140O} found that star-disc interactions influencing the circumstellar discs are more frequent
than previously assumed \citep{2001MNRAS.325..449S}. Not only does the encounter frequency depend on the stellar density, but as
well the prevalent type of encounter. In sparse clusters distant parabolic two-body encounters dominate, whereas in dense
  clusters close encounters often involving several stars at once become increasingly important \citep{2010A&A...509A..63O}.

Observations confirm that the cluster disc fraction (CDF) depends strongly on the stellar cluster density. For example,
\citet{2008ApJ...675.1375L} found significantly more dissolved discs in the dense IC~$348$ cluster compare to the equal-aged, but
sparse Chamaeleon~I cluster. Similarly, \citet{2010ApJ...718..810S} detect a strong decrease of the disc fraction close to the
dense centre of the Arches cluster, one of the densest stellar populations in the Milky Way ($\rho_{\mathrm{core}}>10^5 \Msun $pc
$^{-3}$. Similar results have been obtained for the starburst cluster NGC $3603$ \citep{2004AJ....128..765S} and the Orion Nebula
Cluster \citep[ONC, ][]{1998AJ....116.1816H}. In addition, observed disc frequencies in sparse stellar associations show a
  slower decrease than in denser clusters \citep{2013A&A...549A..15F}.
However, these observations can so far not distinguish whether encounters or photo-evaporation are the dominant environmental
process of disc destruction.

In the following the question will be adressed how the mass distribution in protoplanetary discs influences encounter-induced
losses in young stellar clusters. First, the method and the setup parameters are detailed in Section~\ref{sec:method}. In
Section~\ref{sec:results} the influence of a varying initial disc-mass distribution on the encounter-induced losses in a stellar
cluster is presented. Finally, a discussion will be given in Section~\ref{Discussion} and the results will be summarised in
Section~\ref{sec:summary}.

\section{Method}
\label{sec:method}

First, we simulated the dynamical interactions between all stars in typical stellar clusters for several Myr of cluster
evolution. We use the code \nbody \citep{1963MNRAS.126..223A, 1974A&A....35..237A, 2003gnbs.book.....A}, which has been modified
by adding an encounter tracking routine \citep[see][for more details]{2012ApJ...756..123O}. For each encounter event the encounter
mass ratios and periastron distances are tracked and used to determine the mass loss in the discs. We used the recent results of
\citet{2012A&A...538A..10S} to determine this mass loss for various mass distributions within the discs.

\subsection{Cluster models} 

In the Nbody simulations we choose the masses of the stars in the cluster according to the initial mass function (IMF) given by
\citet{2001MNRAS.322..231K}. Stellar masses below the hydrogen burning limit ( $m_{\mathrm{star}} < 0.08$ $\Msun$) are neglected
since they contribute only little to the stellar dynamics \citep{2005ApJ...625..385A}.

In our simulations all stars are assumed as being initially single. Primordial binaries have been neglected for two reasons:
first, using only single stars reduces the computational cost in the cluster simulations and, second, treating the mass loss in
the disc involving three stars would add another set of variables to the already extensive parameter study of encounters. However,
during the cluster evolution a small fraction of binaries form via capture processes. Further encounters of these systems are
excluded from the disc mass loss \citep{2008IAUS..246...69P}. The effect of primordial binaries on the results for the initial
disc-mass distributions will be discussed in Section~\ref{Discussion}.

The frequncy of star-disc encounters in a young cluster is largely determined by its stellar number density distribution. The ONC
is regarded as the prototype of an embedded dense cluster. Observations of today's ONC density distribution show that it can be
approximated by an isothermal profile ($\propto r^{-2}$) in the outer parts \citep{1994AJ....108.1382M, 1998ApJ...492..540H} and a
flat stellar density profile of the form \mbox{$\rho_{\mathrm{core}} \propto r^{-0.5}$} in the cluster core
\citep{2005MNRAS.358..742S}.
Here, an isothermal density model with an initially slightly increased density in the cluster core region is used, given by

\begin{equation}
  \rho_{\mathrm{initial}}(r) = \begin{cases} \rho_0 \cdot r^{-2.3} &  r \leq R_{\mathrm{core}} \\
    \rho_0 \cdot r^{-2.0}  &  R_{\mathrm{core}} < r \leq R_{\mathrm{cluster}}\\
    \qquad 0 &  R_{\mathrm{cluster}} < r \end{cases}
  \label{olczak_distribution}
\end{equation}
where $\rho_0 = 3.1 \times 10^2 \pc^{-3}$, $R_{\mathrm{core}} = 0.2 \pc$, and $R_{\mathrm{cluster}} = 2.5 \pc$. This distribution develops within \mbox{$1$ Myr}, the estimated age of the ONC, to the stellar distribution observed for the ONC today \citep{2006ApJ...642.1140O}.
Following the approach by \citet{1998MNRAS.295..691B}, here, initially the four most massive stars have been placed in the
inner cluster region with $r_{\mathrm{seg}} = 0.6 \cdot r_{\mathrm{hm}}$, where $r_{\mathrm{hm}}$ is the half-mass radius of the
cluster.

We model clusters of different density by keeping the size of the cluster constant (\mbox{$r_{\mathrm{cluster}} = 2.5 \pc$)} and
varying the initial number of stars $N_{\mathrm{stars}}$. The here covered density range is representative for \textit{massive}
clusters in the solar neighbourhood in the late embedded stage.
Using a slightly steeper profile of $\rho_{\mathrm{core}}(r) \propto r^{-2.3}$ in the cluster core has only a minor effect.

For our simulations of embedded clusters we use the standard method of no gas being included. Since we are in particular
interested in the effect of stellar encounters rather than the cluster evolution the focus here will be on initially virialised
systems of $Q = 0.5$. The stellar velocities are chosen accordingly. To model the stellar velocities, a Maxwellian velocity
distribution with radius-independent velocity dispersion $\sigma$ has been used.  Despite being in virial equilibrium the
  cluster still expands to some degree, so that the stellar density is highest in the earliest stages of the development.

\begin{table}
  \centering
  \begin{tabular}{|c|c|c|cc|}
    \hline
    Model & $N_{\mathrm{stars}}$ & $N_{\mathrm{sim}}$ & $\rho_{\mathrm{centre}} [ 10^3 \pc^{-3}]$ & $\rho_{\mathrm{cluster}}    
    [\pc^{-3}]$ \tabularnewline 
    \hline
    A  &  1\, 000  & 500 &  ~~1.3 & ~~15.3\tabularnewline
    B  &  2\, 000  & 250 &  ~~2.7 & ~~30.6\tabularnewline
    C  &  4\, 000  & 125 &  ~~5.3 & ~~61.1\tabularnewline
    D  &  8\, 000  &  70 &   10.5 & 122.2\tabularnewline
    E  & 16\, 000  &  32 &   21.1 & 244.5\tabularnewline
    F  & 32\, 000  &  16 &   42.0 & 489.2\tabularnewline
    \hline
  \end{tabular}
  \caption{The number of initial stars $N_{\mathrm{stars}}$, number of performed simulations $N_{\mathrm{sim}}$, number density of
    the cluster centre region $\rho_{\mathrm{centre}}$ and total cluster density $\rho_{\mathrm{cluster}} $ for the six investigated
    models are shown. The radius of the core region is \mbox{$r_{\mathrm{centre}} = 0.3 \pc$} while the cluster size is
    \mbox{$r_{\mathrm{cluster}} = 2.5 \pc$.} The core density is given by \mbox{$\rho_{\mathrm{centre}} = 3N_{\mathrm{centre}} / 4 \pi
      r_{\mathrm{centre}}^3$} and the cluster density by \mbox{$\rho_{\mathrm{cluster}} = 3N_{\mathrm{stars}} / 4 \pi
      r_{\mathrm{cluster}}^3$.}}
  \label{tab:parameters__density_scaled_clusters}
\end{table}

It is indispensable to perform multiple simulations ($N_{\mathrm{sim}}$) for each individual cluster model to obtain statistically
robust results. Therefore, a sets of random initial stellar positions, velocities and masses has been achieved for each run, which
are analysed and averaged in a subsequent step. It turned out that tracking at least $500 \, 000$ stellar trajectories
\citep{2013A&A...555A.135P} leads to an error of $< 3 \%$ for each of the presented results. Here, for an initial number of stars
$N_{\mathrm{star}} = 1 \, 000$ around $500$ simulations were preformed while $16$ simulations were performed in case of
$N_{\mathrm{star}} = 32 \, 000$.

An overview of the parameters of the simulated models, like initial number of stars $N_{\mathrm{stars}}$ and corresponding number
of simulations $N_{\mathrm{sim}}$, is shown in Table~\ref{tab:parameters__density_scaled_clusters}, as well as the resulting
densities in the cluster center region $\rho_{\mathrm{centre}}$ (with $r_{\mathrm{centre}} = 0.3 \pc$, e.g. the radius of the
Trapezium cluster) and the mean cluster density $\rho_{\mathrm{cluster}}$ within the cluster radius $r_{\mathrm{cluster}} = 2.5
\pc$.  For details about the dynamical evolution of the simulated clusters see \citet{2010A&A...509A..63O} - e.g. for the
  temporal development of the surface density profile (their Fig.$2$) and the cluster disc fraction (their Fig.$3$).

\subsection{Disc mass loss}

During the cluster simulations for each encounter the mass ratio and the periastron are tracked. Only afterwards, in the diagnostics part, the respective mass losses for all encounter events are determined.
These losses depend on the mass distribution in the disc before the encounter.
Here the tabulated data detailed  in \citep{2012A&A...538A..10S} are used, where first-order interpolations were applied for the values between the tabulated data. These data \citep[see][for more details]{2012A&A...538A..10S} are for coplanar, parabolic encounters. As such they represent upper limits for the disc mass losses.

Initially all stars are surrounded by a disc of size $r_{\mathrm{disc}} = 150 \AU$. This value represents an average value of
observed disc sizes. Here we investigate different power-law mass distributions of the disc material of the form
\begin{equation}
  \Sigma(r) \propto r^{-p} \mathrm{.}
 \label{eq:surface_density_p}
\end{equation}
 This corresponds to the standard form used for fitting in observations.  Observationally determined indices $p$ range roughly from $p = 0$ to $p = 2$ \citep[see][for more details]{2012A&A...538A..10S}. Here we treat specifically the cases $p$ = 0, $p$ = 1, and $p$ = 7/4.

\section{Results}
\label{sec:results}
In the following the focus will be on the central cluster regions ($r_{\mathrm{centre}} = 0.3 \pc$), representing the densest part
of the stellar population.  Here a large fraction of stars is involved in encounter events, which makes it easier to spot the
influence of the different mass distributions.  The simulations cover only the first 2 Myr of cluster development because the
embedded phase of clusters lasts typically 1-3 Myr.

Several studies demonstrated that the typical kind of encounter depends strongly on the cluster density (Olzcak et al. 2010, Dukes
\& Krumholz 2012, Olczak et al. 2012). In low-density clusters gravitational focussing by the massive stars dominates the disc
destruction process, whereas in very dense clusters the interactions between low-mass stars become important.

\subsection{Cluster disc fraction}

\begin{figure}[t]
  \centering
  \includegraphics[width=\linewidth]{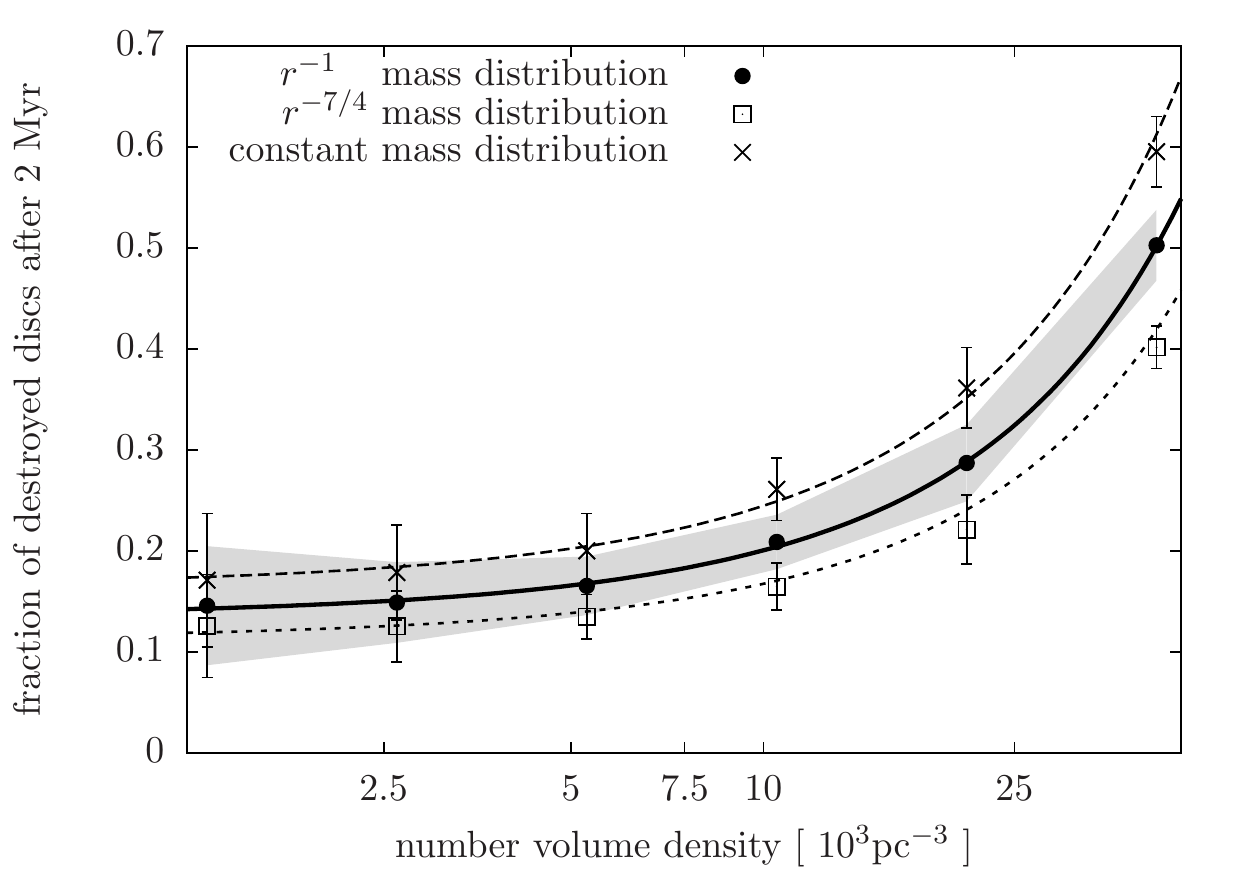}
  \caption{Shown is the fraction of destroyed discs after $2$ Myr of evolution as a function of the stellar number volume density
    in the central cluster region ($r_{\mathrm{center}} = 0.3 \pc$). Circles represent the results of an initial $r^{-1}$
    disc-mass distribution, while crosses show the constant and open squares the $r^{-7/4}$ distributions. The lines represent a
    smooth bezier fit to the data points. The filled curve indicates the standard deviation for $p=1$, which is of the same order
    for the other distributions as indicated by the error bars.}
  \label{fig:encounter_density}
\end{figure}

Figure~\ref{fig:encounter_density} shows the fraction of destroyed discs, $F_{\mathrm{destroyed}}$, after $2$ Myr of cluster
evolution as a function of the central cluster densities in the range $1 \times 10^3$ to $4 \times 10^4 \pc^{-3}$ (represented
by Model \textit{A} - \textit{F}). For initial disc-mass distributions of $r^{-1}$ (filled circles) central cluster densities of $10^3 \pc^{-3}$ lead to
around $15 \%$ of all circumstellar discs being destroyed, while for volume densities above $10^4 \pc^{-3}$ the fraction of
destroyed discs increases considerably to up to $50 \%$ \footnote{The fraction of destroyed discs is slightly lower than in
  \citet{2010A&A...509A..63O}. The reason is that they used a fit formula for the disc-mass losses that slightly overestimates the
  losses in case of mass ratios $M_2/M_1 > 20$. Here, an interpolation algorithm based on an extended parameter study is used.}.  It can be seen that the general trend is the same for different initial mass distributions in
the disc.

Using an initial disc-mass distribution that differs from the $r^{-1}$ case, 
the maximum {\em relative} difference in the fraction of destroyed discs to the standard case
of a $r^{-1}$ distribution is about $25 \%$.
Since in general more discs are tidally destroyed in dense clusters one obtains that the higher the
cluster density the larger are the {\em absolute} differences in the fraction of destroyed discs for the different mass
distributions. Here, an initially constant disc-mass distribution provides a maximum for the encounter-induced
disc mass losses, as material from the outer disc parts is removed.

In other words, for low and intermediate densities, $\rho_{\mathrm{centre}} \leq 10^4 \pc^{-3}$, the cluster disc fraction for the
different initial distributions are found to differ only slightly from the results found for an $r^{-1}$ distribution and are most likely not 
observationally detectable with current number statistics.
However, in the case of $\rho_{centre} = 4 \times 10^4 \pc^{-3}$ (Model \textit{F}) for an initially constant disc-mass distribution the majority
of circumstellar discs ($60 \%$) are destroyed by encounters, whereas for the $r^{-7/4}$ distribution the fraction is only about
$40 \%$. This means that in very dense clusters, where the total fraction of destroyed discs increases significantly, the initial
disc-mass distribution might be important.

\begin{figure}[t]
  \centering
  \includegraphics[width=\linewidth]{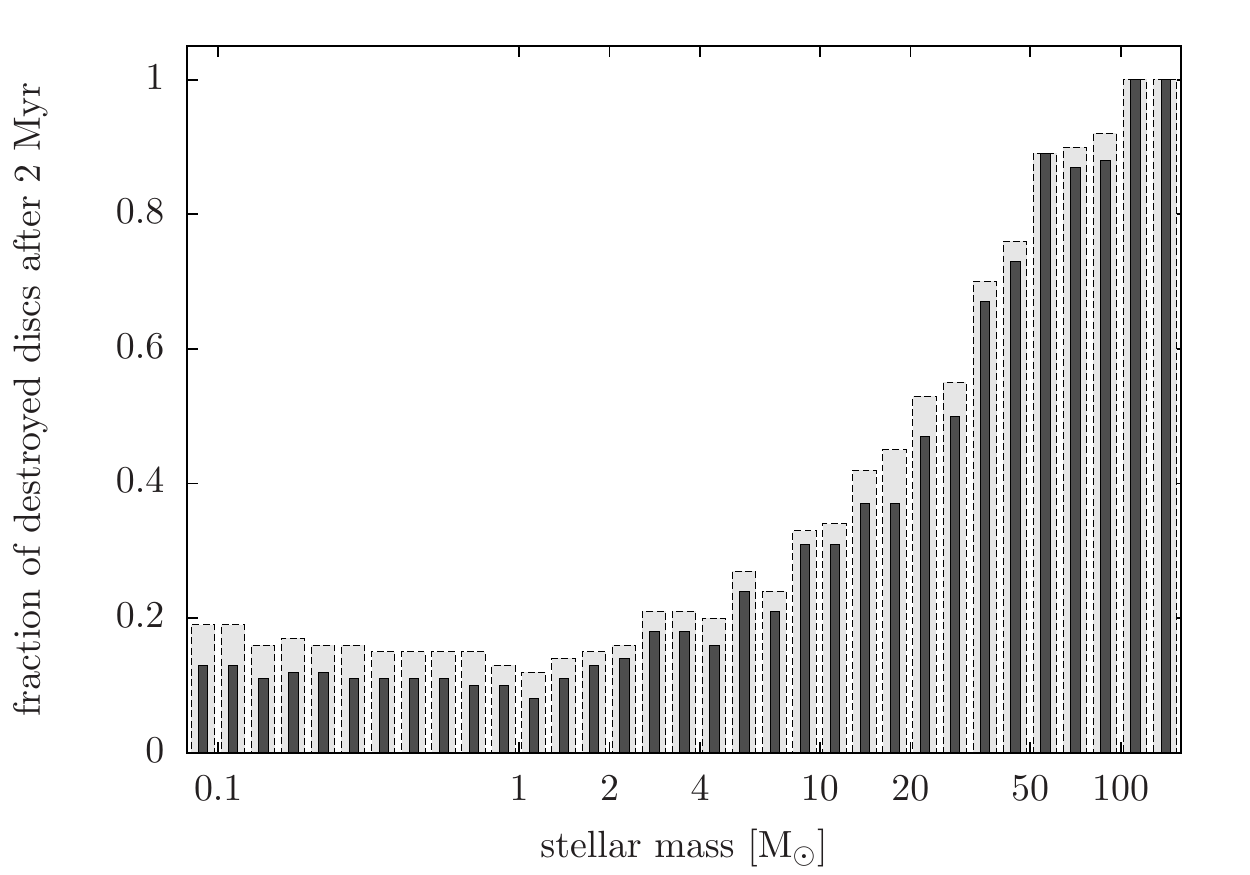}
  \includegraphics[width=\linewidth]{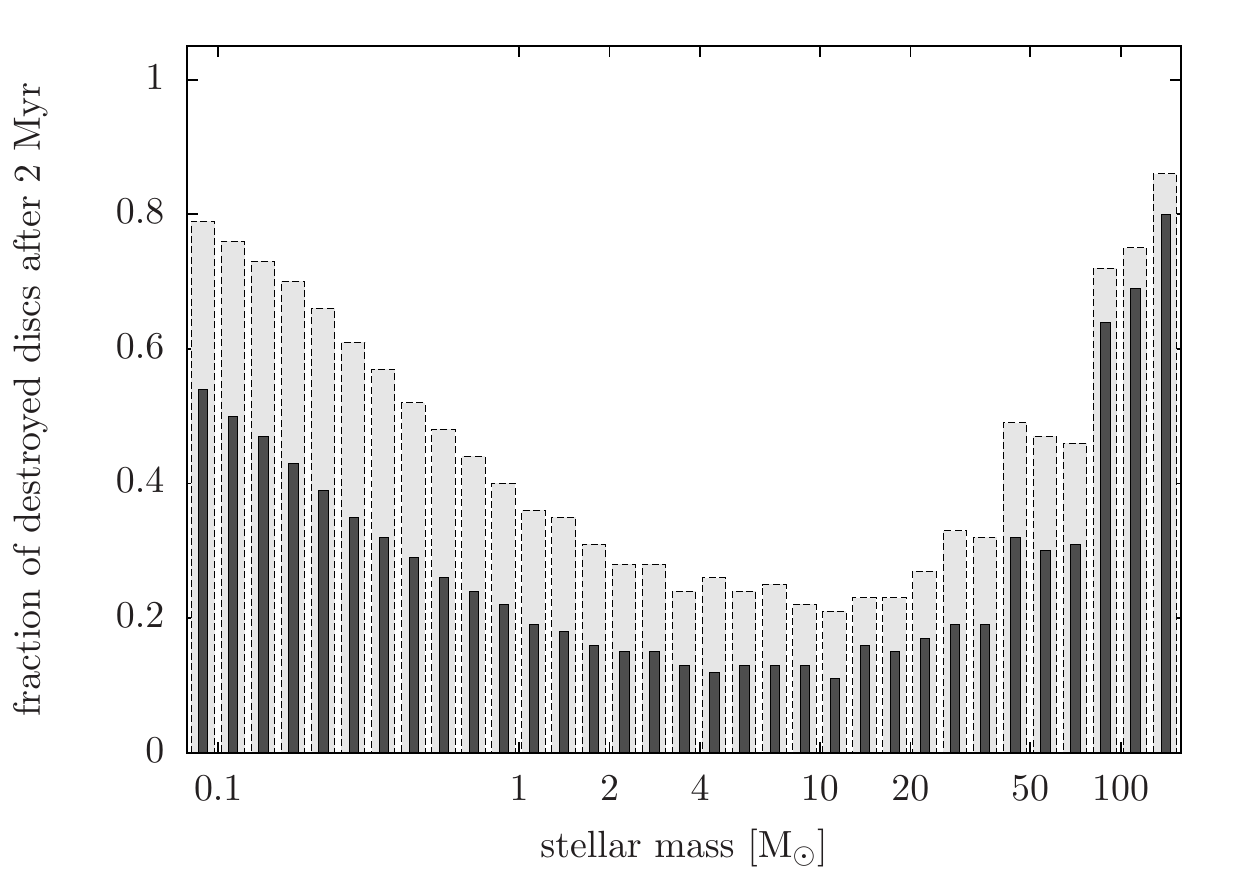}
  \caption{Shown is the fraction of destroyed discs after $2$ Myr as a function of the stellar mass for (a) Model \textit{A} and
    (b) Model \textit{F}. The results for an initially constant disc-mass distribution are shown as grey bars and the $r^{-7/4}$
    case is indicated by black bars. The average deviation per mass bin is $< 5\%$. Stellar masses $> 50 \Msun$ have
    been excluded due to a low number of stars in these mass regime and larger errors of up to $10 \%$.}
  \label{destroyed_disc_vs_mass} 
\end{figure}

How do the differences between the cluster disc fractions depend on the stellar masses involved?
Figure~\ref{destroyed_disc_vs_mass} shows the fraction of destroyed discs as function of the stellar mass for Model~\textit{A}
(Fig.~\ref{destroyed_disc_vs_mass}a) and Model~\textit{F} (Fig.~\ref{destroyed_disc_vs_mass}b). It can be seen that again for
sparser clusters (Fig.~\ref{destroyed_disc_vs_mass}a) differences between an initially constant and an $r^{-7/4}$ disc-mass
distribution on the stellar masses are of minor importance as they are usually well below $5 \%$. For low-mass stars less than $20
\%$ of all stars loose their discs while for high-mass stars the discs are generally completely destroyed by multiple interactions
with the surrounding stars.

For dense clusters (Fig.~\ref{destroyed_disc_vs_mass}b), by contrast, the relevance of the initial disc-mass distribution depends
strongly on the mass of the disc-surrounded star. For stellar masses $M_1 > 2 \Msun$ the differences in the fraction of disc-less
stars usually remain below $\ll 10 \%$.  This is despite the fact that up to $80 \%$ of circumstellar discs are destroyed. The
reason lies in the encounter mass ratios, $M_2/M_1$. Independent of the cluster density, for large stellar masses the median
encounter mass ratio is $M_2 / M_1 \leq 1$. For such low mass ratios the differences between the disc losses for the various
disc-mass distributions is small (see Steinhausen et al. 2012).

By contrast, large encounter mass ratios are obtained for low mass stars $M_1 < 2 \Msun$, where the median encounter mass ratio is
typically $M_2 / M_1 > 10$ in dense clusters. In the case of $M_1 < 2 \Msun$ the differences between the losses for the different
initial disc-mass distributions increase to about $23 \%$. Since most of the cluster members are low-mass stars, the differences
between the total cluster disc fractions is dominated by the fraction of discs around such low-mass stars. As in very dense
clusters, low-mass stars contribute significantly to the disc destruction process, the disc mass overall losses increase while at
the same time the differences between the investigated disc-mass distributions become more pronounced.

\subsection{Disc properties}
\label{Dependence on stellar mass}

In section 3.1 the focus was on strong perturbations that eventually lead to complete disc destruction.  These destructive
encounters are relatively rare events. Much more common are weaker stellar encounters, which do not lead to disc destruction but
can have a significant influence on the disc properties, such as total disc mass, disc size and the mass distribution within the
disc. This process is often overlooked, but it might be of major importance for the properties of the potentially forming
planetary system.

Next we will include these weaker interactions in our investigation to determine their general influence on the disc
properties. We take into account any encounter that leads to $\sim$ 5 \% change in the disc's angular momentum. This particular
value was chosen, because in this case generally the complete disc mass remains bound to the disc-surrounded star, while
nevertheless a significant perturbation of the disc outskirts happens \citep{2003ApJ...592..986P, 2007A&A...462..193P}. In the
following these discs are denoted as perturbed discs.

\begin{figure}[t]
  \centering
  \includegraphics[width=\linewidth]{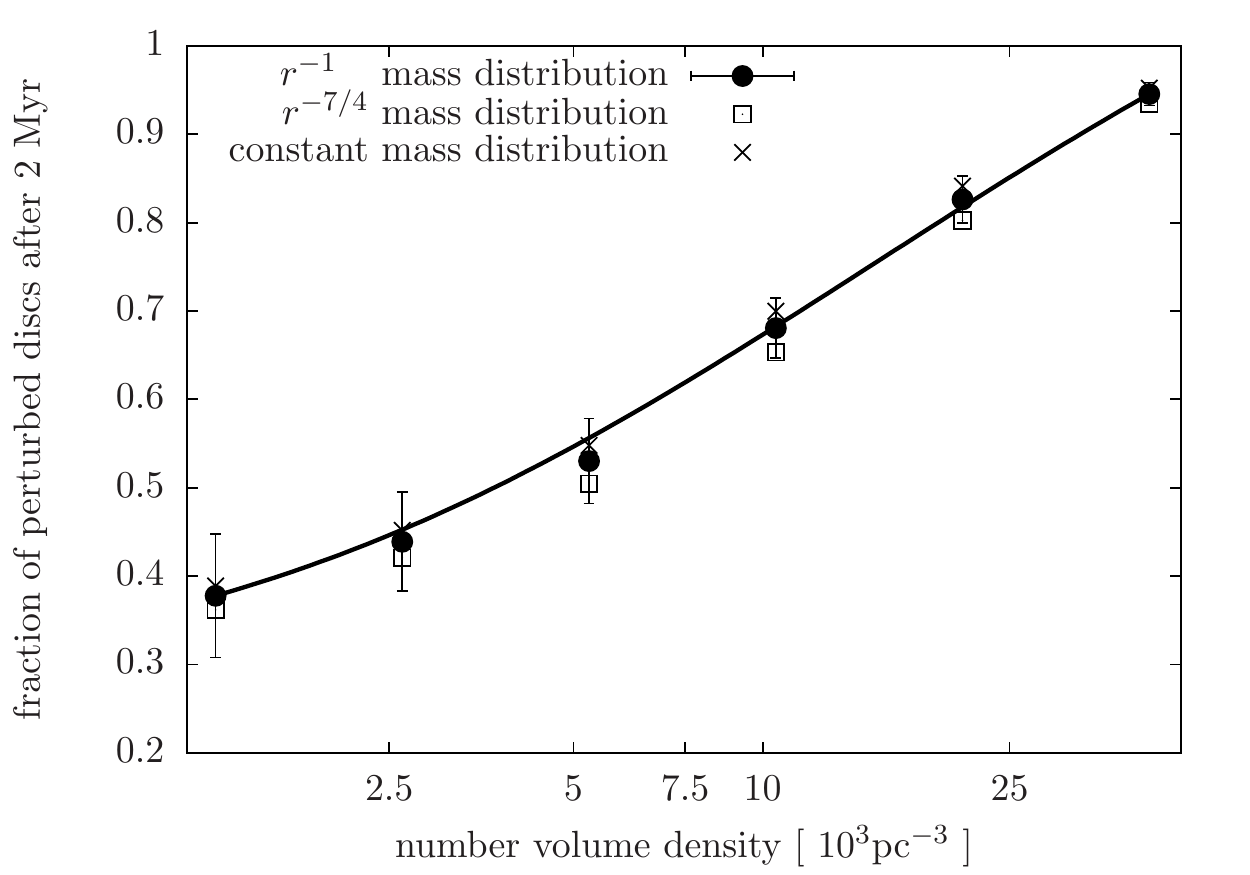}
  \caption{Shown is the fraction of stars with perturbed discs after $2$ Myr of evolution as a function of the stellar number
    volume density in the central cluster region ($r_{\mathrm{centre}} = 0.3 \pc$). Circles represent the results of an initial
    $r^{-1}$ disc-mass distribution, while crosses show the constant and open squares the $r^{-7/4}$ distributions. The line
    represents a smooth bezier fit to the data points for the $r^{-1}$ distribution. The error bars indicate the standard
    deviation for $p=1$, which is of the same order for the other distributions.}
  \label{fig:ang_encounter_density}
\end{figure}

Figure~\ref{fig:ang_encounter_density} shows the fraction of stars with perturbed discs after $2$ Myr as a function of the central
cluster density. We find that the fraction of stars with perturbed discs increases constantly from $40 \%$ for clusters with 
$\rho_{\mathrm{centre}} = 1 \times 10^3$ to $95 \%$ for central densities of $4 \times 10^4 \pc^{-3}$. In dense clusters almost
every star experiences an encounter event within the first $2$ Myr in the dense inner cluster regions.\footnote{Note, that the
  standard deviation for sparse clusters is found to be up to $7 \%$, while the deviations for dense clusters are only $ < 1
  \%$. The reason is that in sparse clusters disc losses are specified by the few high-mass stars and depend strongly on their
  position within the cluster. In dense clusters an increasing number of low-mass stars is involved in encounter events reducing
  the prominent effect of the high-mass stars.}

As weak, distant encounters are much more common than strong ones, one would expect a difference between the different disc-mass
distributions for the number of stars that experience a perturbing encounter. Figure~\ref{fig:ang_encounter_density} shows as
well the results for a constant initial disc-mass distribution (crosses) and a $r^{-7/4}$ distribution (open squares).
Surprisingly, the fraction of stars with perturbed discs is basically the same for all initial disc-mass distributions. The reason
is that in sparse clusters the encounter statistics are dominated by interactions with the massive stars in the cluster centre,
which act as gravitational focii for the low mass stars and usually strongly perturb their stellar discs. By contrast, in dense
clusters multiple interactions between the low-mass stars result in nearly all discs being strongly perturbed. In total, the
differences in the fraction of perturbed discs between the investigated initial disc-mass distributions are minor.

\begin{figure}[t]
  \centering
  \includegraphics[width=\linewidth]{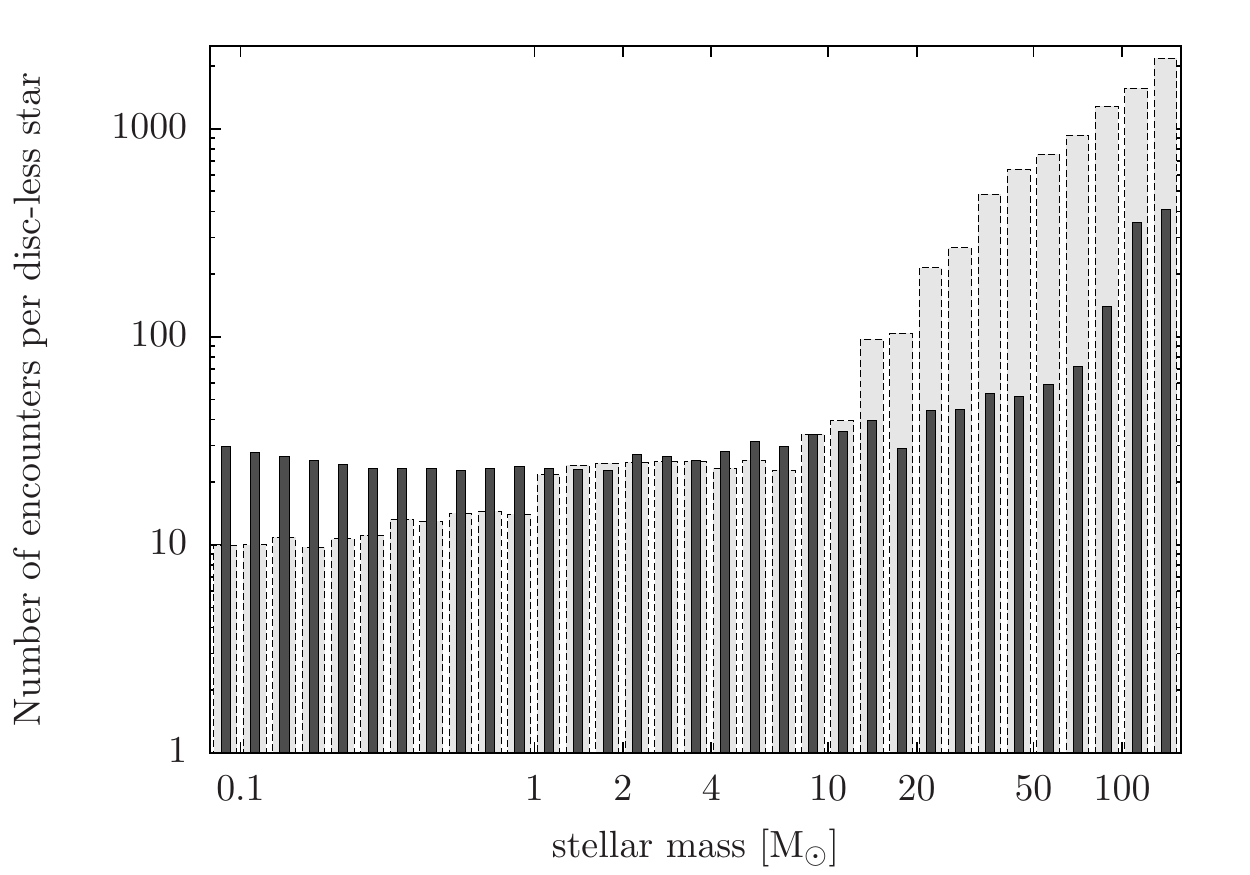}
  \caption{The averaged number of encounters per disc-less star after $2$ Myr of evolution is shown as a function of the mass of
    the central star for Model \textit{A} (dashed gray) and Model \textit{F} (solid black). A constant initial disc-mass
    distribution was used since the number of encounters does only slightly deviate for the different investigated initial
    disc-mass distributions. The average deviation per mass bin is $< 3\%$.}
\label{phd__num_of_enc_vs_mass__1000__2Myr}
\end{figure}

As mentioned before, despite the lower encounter mass ratios in the high-mass star regime the fraction of destroyed discs remains
high due to multiple encounter events. Here again the answer lies in the number of perturbing
encounters. Figure~\ref{phd__num_of_enc_vs_mass__1000__2Myr} shows the averaged number of encounters per finally disc-less star as
a function of the mass of the central star after $2$ Myr for the inner cluster region. Whereas in sparser clusters (Fig. 4, light
grey) low-mass stars experienced usually less than ten encounters per star, this number increases significantly for high-mass
stars. On average the most massive stars ($M_1 > 50 \Msun$) undergo more than $500$ encounters.

For dense clusters, like Model \textit{F} (Fig. 4, black), the situation is different: the number of encounters for low- and
intermediate-mass stars increases significantly. Gravitational focusing becomes less important and the number of encounters for
high-mass stars decreases. The result is a nearly equal number of encounters for low- to intermediate mass stars. Each star
experiences $\sim$ 30 encounter events. The exceptions are the few of the most massive stars, which experience fewer encounters
than in low density clusters but still $\sim$10-times more than the rest of the stars with destroyed discs.

\section{Discussion}
\label{Discussion}

The here presented results can be regarded as upper limits for the influence of encounters on protoplanetary discs in the different stellar environments. One reason is that in above calculations each star was assumed to be initially surrounded by a disc, at the same time only encounters with a
disc-less perturbing star have been investigated. However, \citet{2005ApJ...629..526P} showed that
star-disc encounter results can be generalised to disc-disc encounters as long as there is no significant mass exchange between
the discs. In case of close encounters the discs might be replenished to some extend, which would lead to an overestimate of the
losses in strong perturbing encounters. Most affected by this simplification would be discs with shallow mass distributions, which would be replenished to a larger degree . However, the consequence would be an even smaller difference between the fraction of destroyed discs for the various mass distributions. So our general result, that the disc mass distribution has little effect on the disc fraction, would still hold. 

Another simplification is the focus on prograde, coplanar encounters. If the cluster is not fast rotating, an alignment of the disc and the encounter plane seems rather unlikely.  However, as long as the inclination is not
larger than $45$ degrees, the disc losses due to inclined encounters are only slightly reduced in comparison to a coplanar encounter \citep{2005A&A...437..967P}. Hence, if the
orientations were completely randomly distributed the losses would be overestimated in $75 \%$ of the encounters. This would be the same for any of the considered mass distribution.

Similarly, it has been assumed that the relative losses remain unchanged in a consecutive encounter. \citet{2004ApJ...602..356P}
showed that for equal-mass perturbers a second encounter results not in the same absolute but the same relative losses as in the
first encounter \citep[see also][]{2009DiplomaThesisTackenb}. However, since the perturbations also lead to a steeper surface
  density profile of the remaining discs, our treatment of repeated encounters might lead again to an overestimate of the losses.

Additionally, encounters have been generally treated as parabolic ($\epsilon = 1$) throughout this work. For hyperbolic
encounters, $\epsilon > 1$), due to the shorter interaction times, the total disc losses would drop considerably. In sparse
clusters such hyperbolic encounters can be generally neglected. By contrast, the average eccentricity of the stellar orbits is
higher in case of dense stellar environments, where the massive stars loose their dominating role as encounter partners
\citep{2010A&A...509A..63O}. All mass distributions should be affected to the same degree by this simplification.

Another significant factor for disc losses is the initial disc size. Small disc sizes lead to higher relative periastron distances
$r_{\mathrm{peri}} = r / r_{\mathrm{disc}}$ and therefore lower disc losses in our calculations. In this context, observations
give no clear picture, providing a multitude of observed disc-mass distributions and sizes. Thus, here, the discs are assumed to
have a radius of $r_{\mathrm{disc}} = 150 \AU$, which is a typical observed value. However, a scaling of the disc size with the
mass of the disc-surrounded star by \mbox{$r_{\mathrm{disc}} = 150 \AU \cdot \sqrt{M_1 [\Msun]/\Msun}$}, as it is obtained if a
fixed force of the stars at the discs outer radius is assumed, would be equally likely. This would result in an increased disc
diameter for massive stars ($m_{\mathrm{star}} > 1 \Msun$) while the disc sizes of low-mass stars ($m_{\mathrm{star}} < 1 \Msun$)
are significantly reduced. In the consequence this would lead to a lower fraction of destroyed discs, since the majority of stars
in the cluster is located in the low mass regime.

All these simplifications might lead to overestimating the disc losses. However, some of the applied assumptions potentially lead
to an underestimation.

First of all, sub-stellar objects ($M_{\mathrm{star}} < 0.08 \Msun$) have been excluded in the present study. In general, the mass
ratios in encounters with such low-mass objects are well below $0.1$, which implies that the disc losses would be sufficiently
small ($< 10 \%$). Hence, the effect should be minor for massive stars. If the encounter is non-penetrating the effect can be
neglected even for low-mass stellar objects.

Furthermore, all encounter processes have been treated as two-body encounters. \citet{2001DiplomaThesisUmbreit} showed that
multiple-body encounters result in larger disc losses. However, the effect strongly depends on the mass and periastron distance of
the involved stars. For the present calculations the most destructive encounter has been used to obtain the disc losses, while the
other encounter partners are most likely either distant or less massive, so that their influence on the losses is less
significant.

Primordial binaries have not been included, which might significantly underestimate the destruction rates of stellar discs especially for tight binaries. While it is suggested that up to $100\%$ of all stars \citep[e.g.][and references
therein]{1995MNRAS.277.1491K} might be initially part of a binary system, it remains unclear how their initial periods are
distributed. Assuming the upper limit case of an initially log-uniform period distribution \citep[e.g.][]{2007AJ....134.2272R} a
fraction of $50 \%$ of all stars would have had a companion with a semi-major axis $\leq 100 \AU$. Apart from an increased
destruction rate of the stellar discs in tight binaries, the cluster dynamics are usually influenced by strong few-body
interactions \citep{1975AJ.....80..809H, 1975MNRAS.173..729H}, which potentially leads to underestimating the number of ejections from the
cluster. Hence, primordial binaries might have a non-negligible effect on the disc fractions and stellar dynamics and further
investigations are needed to give an estimate of the effect.

Finally, a large fraction of disc-less stars is ejected after an encounter event within the first few $10^5$ yr with velocities
larger than $20\pc/\mathrm{Myr}$, populating regions of $> 20 \pc$ distance from the cluster centre. In the context of planet
formation such stellar high-velocity escapers from embedded clusters are expected to show no infrared excess emission and to be
less frequently surrounded by planets. Similar results have been obtained from observational \citep{1997AJ....113.1733H,
  2004AJ....128.1254L} and numerical studies that assumed primordial mass segregated populations
\citep{2008A&A...488..191O}. However, such high-velocity escapers are less frequent for primordial non-mass segregated
clusters. Here, a first approach showed that in contrast to evenly distributed clusters, in mass-segregated clusters a lower
fraction of disc-less stars remains in the core region as more stars are ejected due to gravitational focusing of the high-mass
stars. The consequence is an increased fraction of disc-less stars in the core region of primordial non-mass segregated clusters
of up to $10 \%$ in the tested extreme cases.

\section{Summary}
\label{sec:summary}

The focus in this work was on the question, in how far the mass distribution within a protoplanetary disc influences the rate of
tidally destructed discs in typical cluster environments. Stellar clusters spanning a large range of densities have been modelled
and the influence of these different environments on the disc fraction has been investigated.  The main results are:

\begin{enumerate}
\item Surprisingly, even though the initial disc-mass distribution significantly influences individual disc losses induced by
  stellar interactions, the fraction of discs that are completely destroyed by encounters remains fairly unaffected, as long as
  the cluster density does not exceed $\rho_{\mathrm{core}} < 10^4 \pc^{-3}$. These are still quite dense clusters, for
  example the ONC would belong to this group.  The reason is that in these clusters the complete destruction of discs happens by
  interactions with high-mass stars. These type of encounter is rather insensitive to the initial disc-mass distribution.  
\item By contrast, for very dense clusters, an example would be NGC 3603, the fraction of destroyed discs depends to some degree
  on the initial disc mass distribution. More specific, $60 \%$ of discs are destroyed assuming initially constant disc-mass
  distributions while for an initially steep disc-mass distribution ($\Sigma \propto r^{-7/4}$) only 40\% are affected. In such
  dense clusters, interactions between low-mass stars are not only more frequent in absolute but as well in relative terms. Such
  encounters between low-mass stars show the strongest dependence on the mass distribution in the disc. However, in general even
  in this case the fraction of destroyed discs deviates no more than $\sim$ 20\% from an initially $r^{-1}$ disc-mass
  distribution.
\item The initial disc-mass distribution has little influence on the total number of stars, that have a disc that is changed in
  its structure by a fly-by. However, independently of the initial distribution of the disc material, almost all circumstellar
  discs (95\%) in the core region ($r_{core}$) of dense clusters are significantly effected by fly-bys.
\end{enumerate}

This means the most results usually obtained assuming a $r^{-1}$ disc mass distribution can be largely generalised to other mass
distributions. The exception might be clusters with densities in excess of $\rho_{\mathrm{core}} > 10^4 \pc^{-3}$ like for example
NGC 3603.





\end{document}